\documentclass[12pt,english,aps,amssymb,showpacs,superscriptaddress,secnumarabic]{revtex4}

\usepackage[T1]{fontenc}
\usepackage[latin1]{inputenc}
\usepackage{float}
\usepackage{graphicx,latexsym,t1enc}
\usepackage{babel}

\begin{document}
\date{\today{}}
\title{Nanoscale sliding friction versus commensuration ratio}

\author{Evy Salcedo Torres} \email{esalcedo@if.ufrgs.br}
\affiliation{Instituto de Física - Universidade Federal do Rio Grande do Sul,
Caixa Postal 15051, 90501-970 Porto Alegre RS, Brasil}

\author{Sebastián Gonçalves} \email{sgonc@if.ufrgs.br}
\affiliation{Instituto de Física - Universidade Federal do Rio Grande do Sul,
Caixa Postal 15051, 90501-970 Porto Alegre RS, Brasil}
\affiliation{Consortium of the Americas for Interdisciplinary Science and\\
Department of Physics and Astronomy - University of New Mexico, Albuquerque, 
New Mexico 87131}

\author{Claudio Scherer} \email{cscherer@if.ufrgs.br}
\affiliation{Instituto de Física - Universidade Federal do Rio Grande do Sul,
Caixa Postal 15051, 90501-970 Porto Alegre RS, Brasil}

\author{Miguel Kiwi} \email{mkiwi@puc.cl}
\affiliation{Facultad de Física - Pontificia Universidad Católica,
Casilla 306, Santiago, CHILE 6904411}

\begin{abstract}  
  The pioneer work of Krim and Widom unveiled the origin of the
  viscous nature of friction at the atomic scale. This generated
  extensive experimental and theoretical activity.  However,
  fundamental questions remain open like the relation between sliding
  friction and the topology of the substrate, as well as the
  dependence on the temperature of the contact surface. Here we
  present results, obtained using molecular dynamics, for the phononic
  friction coefficient ($\eta_{ph}$) for a one dimensional model of an
  adsorbate-substrate interface.  Different commensuration relations
  between adsorbate and substrate are investigated as well as the
  temperature dependence of $\eta_{ph}$. In all the cases we studied
  $\eta_{ph}$ depends quadratically on the substrate corrugation
  amplitude, but is a non-trivial function of the commensuration ratio
  between substrate and adsorbate.  The most striking result is a deep
  and wide region of small values of $\eta_{ph}$ for
  substrate-adsorbate commensuration ratios between $\approx 0.6-0.9$.
  Our results shed some light on contradictory results for the
  relative size of phononic and electronic friction found in the
  literature.
\end{abstract}

 \maketitle

\section{Introduction}
\label{int}

The pioneer work of Krim and Widom, revealing the viscous nature of
the friction of a krypton monolayer sliding over a gold
substrate~\cite{Krim88}, generated a flurry of theoretical work
intended to develop an understanding of this interesting
phenomenon~\cite{Smith,Persson96,Tomassone97,Liebsch99}.
Subsequently, a huge body of new and fascinating experiments on
nanoscale friction have seen the light during the last 15 years.
However, the theoretical interpretation of nanoscopic sliding
experiments has evidenced some degree of disagreement, mainly related
to incompatible results between different
simulations~\cite{Smith,Persson96,Tomassone97,Liebsch99}, in spite of
the fact that they were performed with similar models and techniques.
With the aim of finding some plausible explanation for these
discrepancies we set out to study the relation between the topology
and the sliding friction coefficient $\eta$. Specifically, we conduct
a careful study of a one-dimensional model to disclose the relation
between sliding atomic friction and substrate-adsorbate commensuration
ratio.

Understanding the origin of sliding friction is a fascinating and
challenging enterprise~\cite{Rob_mus,Mus_rob}. Issues like how the
energy dissipates on the substrate, which is the main dissipation
channel (electronic or phononic) and how the phononic sliding friction
coefficient depends on the corrugation amplitude were addressed, and
partially solved, by several
groups~\cite{Smith,Persson96,Tomassone97,Liebsch99}.  Yet, in the
specific case of Xe over Ag, four different groups tried either to
explain, or to estimate theoretically, the experimental observations
of Daily and Krim~\cite{Daly96}. Despite the fact that they used
similar models, and similar simulation techniques to calculate the
phononic contribution to friction, they got quite different results.
Persson and Nitzan~\cite{Persson96} found that the phononic friction
was not significant in comparison with the electronic contribution,
while Smith et al.~\cite{Smith}, and Tomassone et
al.~\cite{Tomassone97} found the opposite. On the other side Liebsch
et al.~\cite{Liebsch99} concluded that both contributions are
important, but that the phononic friction strongly depends on the
substrate corrugation amplitude.  The abrupt change in the sliding
friction at the superconductor transition observed by Dayo et
al.~\cite{Dayo98} provides additional support to the latter argument,
showing that the electronic friction is of the same order of magnitude
as the phononic one.

However, this relation between corrugation amplitude and phononic friction
cannot explain the full magnitude of the disagreement between the different
authors. To achieve full agreement we would be forced to allow for huge
differences in the corrugation amplitude between the various models, much larger
than in actual fact.  An apparently subtle technical detail or artifact of the
molecular dynamics simulation might show the path to the answer of such
divergences, either by itself or by clarifying and improving on the above
mentioned corrugation dependence.

When carrying out molecular dynamics simulation the Xe adsorbate adopts one of
two different orientations relative to the 110 Ag substrate.  This in turn
produces significant changes in the effective sliding friction coefficient,
which maybe due to the different commensuration ratios between substrate and
adsorbate, since the adsorbate adopts one or the other of the two preferred
orientations.  With the objective of elucidating this question here we present
molecular dynamics results for a one dimensional system. Our aim is to stress
the role, and at the same time to have complete control of, the relation between
$\eta$ and the commensuration ratio.

\section{Model}
\label{mod}
\begin{figure}[htb]
\center
\includegraphics[width=7cm]{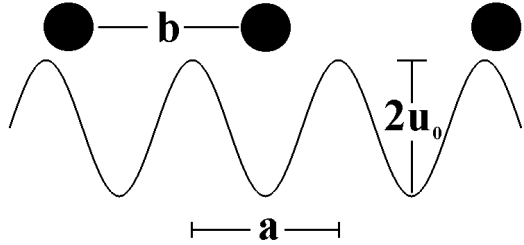}
\caption{The one-dimensional model}
\label{model}
\end{figure}

The model, schematically depicted in Fig.~\ref{model}, can be thought
of as a generalized Frenkel-Kontorova model plus a fluctuation-dissipation
mechanism.  It consists of a one-dimensional chain of atoms that interact with
each other through a Lennard-Jones interatomic potential (adsorbate), moving in
a periodic external potential (substrate).  Apart from the interatomic and the
adsorbate-substrate interaction, there are damping and stochastic forces that
act as a thermostat, plus an external applied force which makes the adsorbate
slide over the substrate. This way, we intend to model the sliding of a solid
monolayer over a perfect crystalline substrate. Therefore, the adatoms of mass
$m$ and labeled by the indices $i$ and $j$, obey the following Langevin
equation:

\begin{equation}
m\ddot{x_{i}}+m\eta_{e}\dot{x_{i}}=-\sum_{j}\frac{\partial
V(\left|x_{i}-x_{j}\right|)}{\partial x_{i}}-\frac{\partial U(x_{i})}{\partial
x_{i}}+f_{i}+F \; .
\label{eq5}
\end{equation}

\noindent 
The external force $F$ is applied to every atom in the chain ---the total
external force is thus $NF$--- and $f_i$ is a stochastic fluctuating force drawn
from a Gaussian distribution, and related to $\eta_e$ via the
fluctuation-dissipation theorem

\begin{equation}
\left\langle f_{i}(t)\; f_{j}(t)\right\rangle
=2\eta_{e} mk_{B}T\delta_{ij}\delta(t)\; ,
\label{eq5-1}
\end{equation}
where $T$ is the temperature of the substrate and $k_{B}$ is the Boltzmann
constant. The stochastic force plus the dissipation term provide a thermal bath
to describe the otherwise frozen substrate.  Moreover, the damping term
represents the electronic part of the microscopic friction, which cannot be
included in a first principles way in our classical treatment.  $\eta_{e}$ may
thus be regarded as the electronic sliding friction coefficient.  The expression
of the interatomic Lennard-Jones potential between adatoms, $V(r)$, where
$r=|x_{i}-x_{j}|$, is given by

\begin{equation}
V(r)=\varepsilon\left[\left(\frac{r_{0}}{r}\right)^{12}-2
\left(\frac{r_{0}}{r}\right)^{6}\right],
\label{eq1}
\end{equation}
where $r_{0}$ is the $T=0$ equilibrium distance of the dimer and $\varepsilon$
is the depth of the potential well. The interaction is cut-off beyond third
neighbors, and hence the cell parameter for the isolated chain is $b=0.9972
r_0$. The adsorbate-substrate potential $U(x)$ is a periodic potential 

\begin{equation}
U(x)=u_{0}\left[\cos\left(2\pi\frac{x}{a}\right)+1\right],
\label{eq2}
\end{equation}
and therefore $a$ is the periodicity of the potential, representing the distance
between neighboring substrate atoms; $u_{0}$ is the semi-amplitude of the
adsorbate-substrate potential, and it is usually called the substrate
corrugation.

In what follows we take $r_{0}$, $\varepsilon$ and the mass $m$ of the adsorbate
atoms as the fundamental units of the problem, expressing all other quantities
in terms of them.  For example, the time unit is
$t_{0}=r_{0}\sqrt{m/\varepsilon}$, and consequently in units of
$r_{0}=\varepsilon = m=1$ it follows that $ t_0 = 1$. As for temperatures they
are presented as $k_B T$, therefore they are in units of $\varepsilon$.

The periodic boundary conditions impose the following relation between $a$ and
$b$:

\begin{equation}
N_{s}a=N_{a}b,
\label{eq3}
\end{equation}
where $N_{s}$ is the number of substrate atoms and $N_{a}$ the number of
adsorbed atoms. Therefore, the commensuration ratio between substrate and
adsorbate is given by

\begin{equation}
\frac{a}{b}=\frac{N_{a}}{N_{s}} \; .
\label{eq4}
\end{equation}
As was stated the Sec.~\ref{int}, several
authors~\cite{Smith,Persson96,Tomassone97,Liebsch99} have already stressed the
important role of coverage and corrugation in the understanding of atomic
sliding friction.  Here we focus on the effect of these parameters, restricting
the system to be one-dimensional, in order to avoid any possible topological
artifacts that are due to finite size effects.  Consequently $a/b$ and $u_0$ are
our key parameters in the study of the adsorbate-substrate interface response to
sliding friction.

Eq.~(\ref{eq5}) is numerically integrated using a Langevin molecular dynamics
algorithm~\cite{Allen80,Allen82,Frenkel} for 2500 particles ($N_a$), with a time
step $\Delta t = 0.01$.  To obtain the friction coefficient $\eta$ an external
force $F$ is applied; however, before this force is applied we allow the system
to relax during $t_r=2500$. Next, $F$ is applied to every adatom, and after
another transient period of the same extension the adatoms are presumed to have
reached the steady state, with the average friction force equal to the external
force (this conjecture is shown below to be valid). Analytically,

\begin{equation}
 F = m \eta v\; ,
\label{force}
\end{equation}
and hence $\eta$ is the effective microscopic friction coefficient which
includes the \textit{ad hoc} friction coefficient $\eta_{e}$ of Eq.~(\ref{eq5}).
In all the calculations presented here we keep the electronic contribution to
$\eta$ fixed at $\eta_e = 5 × 10^{-3}$; as the effective friction coefficient is
$\eta = \eta_e + \eta_{ph}$, the precise value of $\eta_e$ is 
irrelevant for the calculation of $\eta_{ph}$.

\begin{figure}[htb]
\center \includegraphics[clip=true, width=7cm]{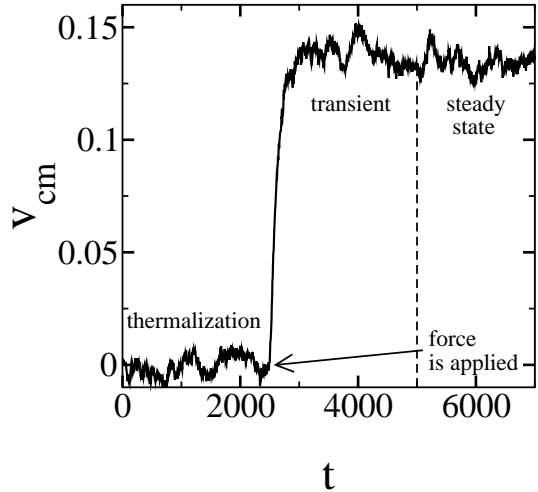}
\caption{Typical behavior of the adatoms center of mass velocity ($v_{cm}$)
during a Langevin molecular dynamics run, before and after the force is applied.
In this case $u_0=0.04\varepsilon$ and $a/b=0.95$. The applied external force is
$F=10^{-3}\varepsilon/r_0$ and temperature is $T=0.05\varepsilon/k_B$. $v_{cm}$
and $t$ in reduced units of $\sqrt{\varepsilon/m}$ and $r_0\sqrt{m/\varepsilon}$
respectively.}
\label{vcm}
\end{figure}

Figure~\ref{vcm} shows the typical behavior of the adatoms center of mass
velocity during a Langevin molecular dynamics run, before and after the external
force is applied. We observe that the system relaxes in a typical time which is
less than $1000$ during the thermalization period, and that it enters into a
steady state in about the same time after the force is applied. The temperature
is $T=0.05$ and the applied force is $F=0.001$. Therefore, in this case, the
resulting friction coefficient is $\eta=0.0075$, which implies
$\eta_{ph}=0.0025$.

\section{Results}\label{res}
As was stated above, in this contribution our main concern is the relation
between atomic sliding friction and adsorbate-substrate interface
topology. Consequently, in this section we present results for our model of the
effective sliding friction coefficient $\eta$, for different values of the
substrate-adsorbate commensuration ratio $a/b$, and of the substrate corrugation
$u_0$. We also check on the effect of substrate temperature.

\begin{figure}[htb]
\center \includegraphics[clip=true,width=7cm]{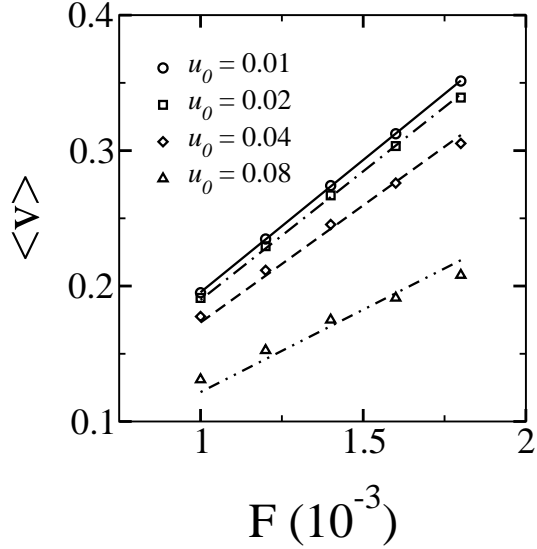}
\caption{Center of mass velocity $v$ as a function of the external force $F$, in
reduced units $[v] = \sqrt{\varepsilon/m}$ and $[F] = \varepsilon/r_0$, for four
different substrate corrugation values and for a commensuration ratio
$a/b=0.451$. The resulting $\eta$ coefficients (in units of
$\sqrt{\varepsilon/(mr_0^2)}$) are: $5.12 \times 10^{-3} (\bigcirc), 5.27 \times
10^{-3} (\Box), 5.78 \times 10^{-3} (\Diamond)$, and $8.22 \times 10^{-3}
(\bigtriangleup)$.}
\label{vxf}
\end{figure}

Once we have made sure that the atomic friction obeys
Eq.~(\ref{force}) we obtain from it the effective friction coefficient
$\eta$ simply by evaluating the slope of the $F\propto v$ relation.
The force $F$ is the input value and the velocity is obtained by time
averaging the steady state center of mass velocity, as exemplified for
one run in Fig.~\ref{vcm}. To obtain reliable data we perform five
independent runs for each value of $F$, and we 
take the average of the resulting $v_{cm}$ values which is denoted as
$\langle v_{cm} \rangle $.  A set of results for $a/b = 0.451$ is
shown in Fig~\ref{vxf}, where it is observed that, although some
deviations are present, the linear relation between $\langle v_{cm}
\rangle $ and $F$ is a valid assumption for all the substrate
corrugation values considered here.  For other commensuration ratios
similar pictures are also obtained.

\begin{figure}[htb]
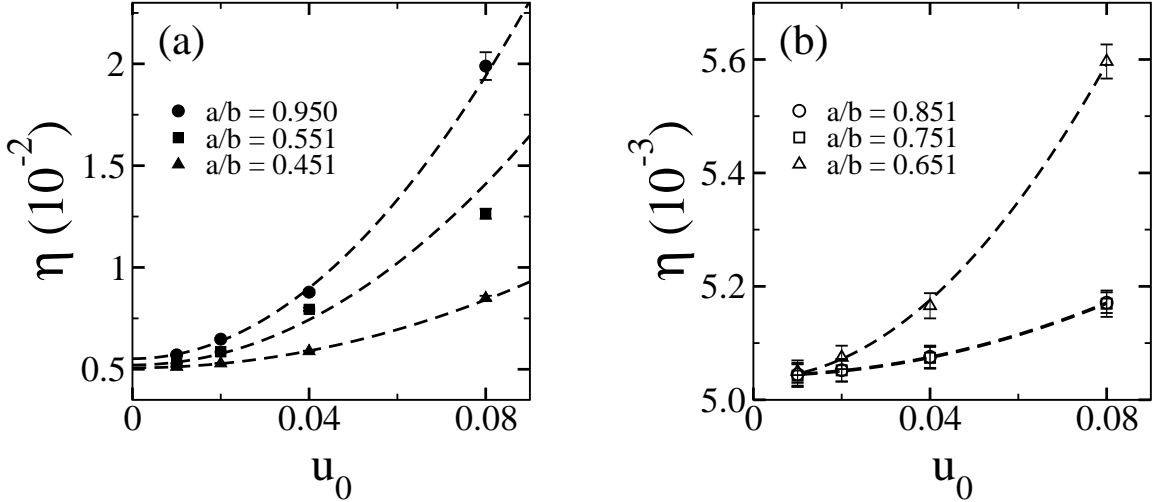

\center \includegraphics[clip=true,width=7cm]{fig4a.eps}
\hspace{1cm} \includegraphics[clip=true,width=7cm]{fig4b.eps}
\caption{Total friction coefficient (in units of $\sqrt{\varepsilon/(mr_0^2)}$)
  for a variety of commensuration ratios $a/b$ and for a temperature
  $T=0.05\varepsilon/k_B$. (a) ``strong friction'' coefficients; (b) ``weak
  friction'' coefficients.}
\label{eta}
\end{figure}

In Figs.~\ref{eta} we present the results for the effective friction coefficient
at $T=0.05$, and for different commensuration ratios. They are separated
into two plots, depending on the range of variation of $\eta$ as a function of
the corrugation amplitude.  Cases where the coefficient $\eta$ rises above $0.9
× 10^{-2}$ at the maximum corrugation amplitude ($u_0 = 0.08$) are shown in
Fig.~\ref{eta}(a), while the cases with friction coefficient below that value (at
the same corrugation amplitude) are displayed in Fig.~\ref{eta}(b). We denominate them,
respectively, ``strong'' and ``weak friction'' cases.

Given the theoretical arguments presented by Smith et al.~\cite{Smith} and
the numerical evidence presented of Liebsch et al.~\cite{Liebsch99}, supporting
the quadratic dependence of the friction coefficient with substrate corrugation,
we fitted the data of Fig.~\ref{eta} with the following expression:

\begin{equation}
\eta = \eta_{0} + c u_{0}^{2},
\label{eq6}
\end{equation}
where $\eta_{0}$ is the friction coefficient in the absence of corrugation, and
thus equal to the ``ad-hoc'' (or electronic) friction coefficient $\eta_e$.  The
second term of the effective friction coefficient, $c u_{0}^{2}$, is referred to
as the phononic friction coefficient, $\eta_{ph}$, which depends quadratically
on the corrugation amplitude $u_0$.  The resulting parabolas can be seen in
Fig.~\ref{eta} (dashed lines)
along with the fitted data, while the coefficients $\eta_{0}$ and $c$
are given in Table~\ref{table2}.  Notice that our results are
consistent with Eq.~(\ref{eq6}), and that $\eta_{0}$ is indeed equal
to $\eta_{e}$.

\begin{table}
\caption{Fitting parameters $\eta_0$ and $c$ of Fig.~\ref{eta}\label{table2}}
\begin{center}
\begin{tabular}{ccr@{$±$}lr@{$±$}l}

\hline
group      & $a/b$ &  \multicolumn{2}{c}{$\eta_{0}\;(10^{-3})$} & 
\multicolumn{2}{c}{$c\;(10^{-2})$} \\
\hline \hline 
``strong'' & 0.451 & 5.07 & 0.02 & 52  & 1   \\
           & 0.551 & 5.21 & 0.02 & 139 & 3   \\
           & 0.950 & 5.52 & 0.03 & 217 & 6   \\
``weak''   & 0.651 & 5.04 & 0.02 & 8.6 & 0.5 \\
           & 0.751 & 5.04 & 0.01 & 2.0 & 0.4 \\
           & 0.851 & 5.04 & 0.01 & 2.0 & 0.4 \\
\hline

\end{tabular}
\end{center}
\end{table}

\begin{figure}[htb]
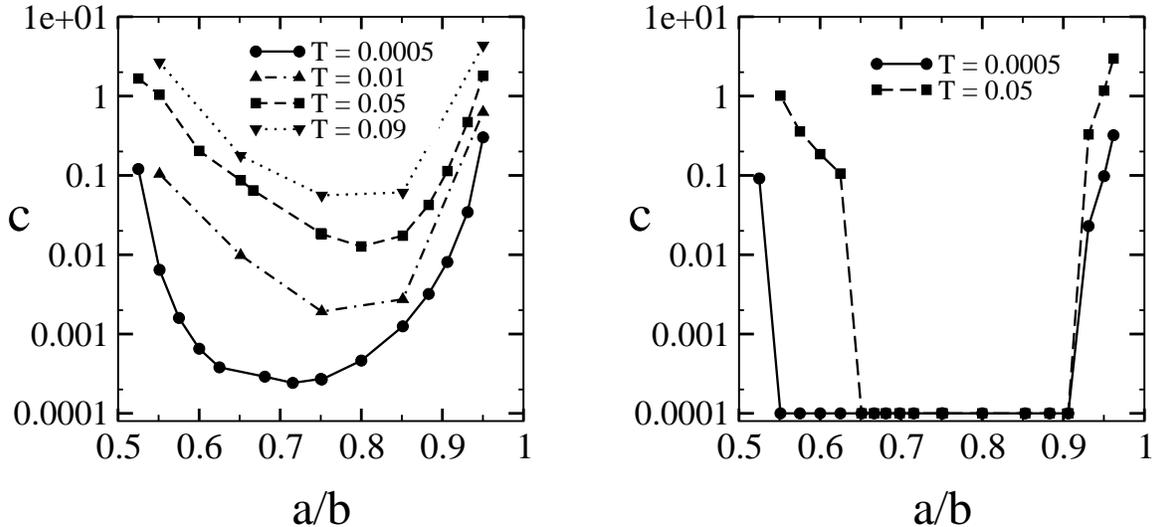

\center \includegraphics[clip=true,width=7cm]{fig5a.eps}
\hspace{1cm} \includegraphics[clip=true,width=7cm]{fig5b.eps}
\caption{Coefficient $c$ of the phononic friction ($\eta_{ph} = c u_0^2$) for
several commensuration ratios. (a) Results obtained with external force $F$ in
the range $0.001-0.002\varepsilon/r_0$ and four different temperatures. (b)
Results obtained with external force $F$ in the range
$0.0001-0.001\varepsilon/r_0$ and two different temperatures. The coefficient
$c$ is in units of $(mr_0^2\varepsilon^3)^{-1/2}$ and temperature is in units of
$\varepsilon/k_B$.}
\label{coef}
\end{figure}

Once it is properly established that the sliding friction (of
adsorbate-substrate interfaces) due to phonons depends quadratically
on the substrate corrugation amplitude, the coefficient $c$ becomes
the main source of information about the phononic friction between
different surfaces.  A synthesis of the results of the present
contribution are given in Fig.~\ref{coef}, where we illustrate the
behavior of the coefficient $c$ of Eq.~(\ref{eq6}) over the whole
range of commensuration ratios, and for several different temperature
values, using a semi-logarithmic plot.  The principal feature of this
figure is the region of low $\eta_{ph}$ values for $a/b \approx
0.6-0.9$. Fig.~\ref{coef}(a) and Fig.~\ref{coef}(b) differ on the
range of forces used in each one of them.  Fig.~\ref{coef}(a) was
obtained using forces in the range $0.001-0.002$, while for
Fig.~\ref{coef}(b) we used forces in the range $0.0001-0.001$.  That
the two set of curves are not identical means that the linear
assumption for the $f$---$v$ relation is valid only if restricted to
relatively small force ranges --less than a decade, as was the case
in all previous contributions on sliding friction.  We will
comment on this below, at the end of this section.

\begin{figure}[htb]
\center \includegraphics[clip=true,width=7cm]{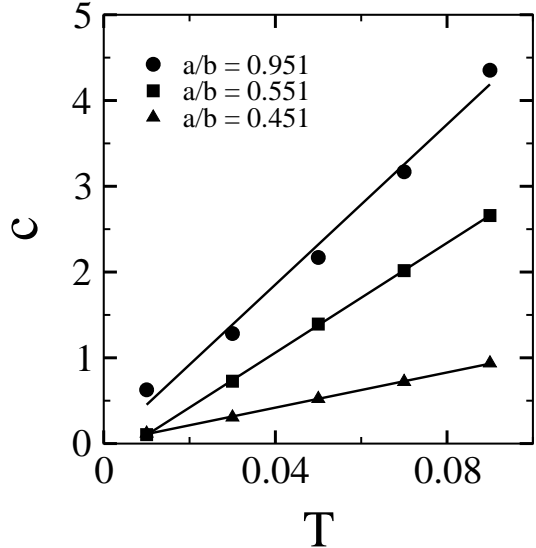}
\caption{Temperature dependence of the coefficient $c$ for several strong
friction commensuration ratio values. The coefficient $c$ is in units of
$(mr_0^2\varepsilon^3)^{-1/2}$ and temperature is in units of
$\varepsilon/k_B$.}
\label{cxt}
\end{figure}

We conclude from Fig.~\ref{coef} that the coefficient $c$ has a non-trivial
relation with the commensuration ratio, which allows to discriminate between
strong and weak phononic friction, {\it i.e.} as a function of interface
mismatch.  Besides, it reflects how the friction coefficient varies with
temperature; to illustrate the latter behavior in detail we present in
Fig.~\ref{cxt} the relation between $c$ and temperature, for some selected
(``strong'' cases) values of the commensuration ratio.  We see that for a fixed
$a/b$ ratio the $c$ coefficient increases linearly with temperature.

\begin{figure}[htb]
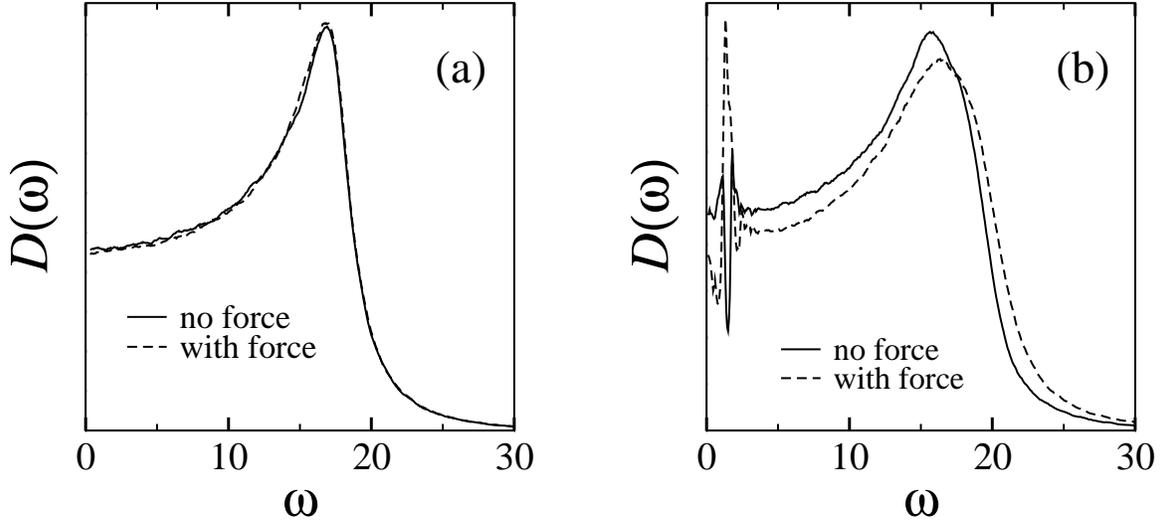

\center \includegraphics[clip=true,width=7cm]{fig7a.eps}
\hspace{1cm} \includegraphics[clip=true,width=7cm]{fig7b.eps}
\caption{Normalized phonon-frequency spectrum for two different type of
  commensuration ratio values, one of the (a) weak friction type ($a/b=0.851$)
  and the other of the (b) strong friction type ($a/b=0.950$) The frequency
  $\omega$ is presented in units of $\sqrt{\varepsilon/(mr_0^2)}$.}
\label{dos}
\end{figure}

In order to obtain some insight on the origin of the sliding friction
``wide minimum'' as a function of commensuration ratio, we look at the adsorbate
phonon density of states ${\cal D}(\omega)$, which can be readily computed, in
a molecular dynamics simulation, by means of the Fourier transform of the
velocity autocorrelation function $\gamma$, given by~\cite{Dickey69}

\begin{equation}
\gamma(t)=\frac{\sum v_{i}(t)\cdot v_{i}(0)}{\sum v_{i}(0)^{2}} \; .
\end{equation}
The resulting ${\cal D}(\omega)$ spectra are shown in Figs.~\ref{dos} for one
representative $a/b$ ratio of each group, both at the low temperature $T=0.005$;
Fig.~\ref{dos}(a) corresponds to the weak and Fig.~\ref{dos}(b) to the strong
friction group. We do so for the $F\neq 0$ (dashed line) and $F=0$ (solid line)
steady state cases, where $F$ is the external applied force.  Notice that for
the strong friction case, extrema of ${\cal D}(\omega)$ in the low frequency
region are generated, which closely resemble one-dimensional Van Hove
singularities; {\it i.e.} points which correspond to extrema (maxima or minima)
of the $\omega$ versus $k$ dispersion relation.  In our model they are due to
the mismatch of the chain and periodic potential periodicities.  When the
external force is applied an inversion occurs: in the frequency region where
there is a minimum for the $F=0$ pre-force situation, a maximum of ${\cal
D}(\omega)$ develops during chain sliding.  This feature is absent in the weak
friction case, where the phonon density of states is much like the typical one
for an isolated 1D system, with no difference between the force free ($F=0$) and
the driven ($F\neq 0$) cases.  As the commensuration ratio $a/b$ crosses over
from the strong to the weak friction regime the maxima are quenched.

The physical implications of the ${\cal D}(\omega)$ maximum are that in a
non-commensurate state some low frequency (large wavelength) modes cannot be
excited, since the adsorbate has to adjust to the mismatch condition. However,
when the system starts its sliding motion, precisely these `soft' modes are the
ones that are excited preferentially, with the consequent amplitude increase in
that frequency region. It is just this feature that singles out the most
efficient energy dissipation channel. The way in which this is achieved is by
pumping center of mass energy into the adsorbate vibration modes that lie in the
vicinity of the Van Hove like singularity, which increases the friction relative
to the weak friction case, where the mismatch does not impose on the system a
frequency region where it does not vibrate `comfortably'.  Therefore, in the
weak friction regime, the phononic dissipation channel is just the usual one.

\begin{figure}
\center \includegraphics[clip=true,width=7cm]{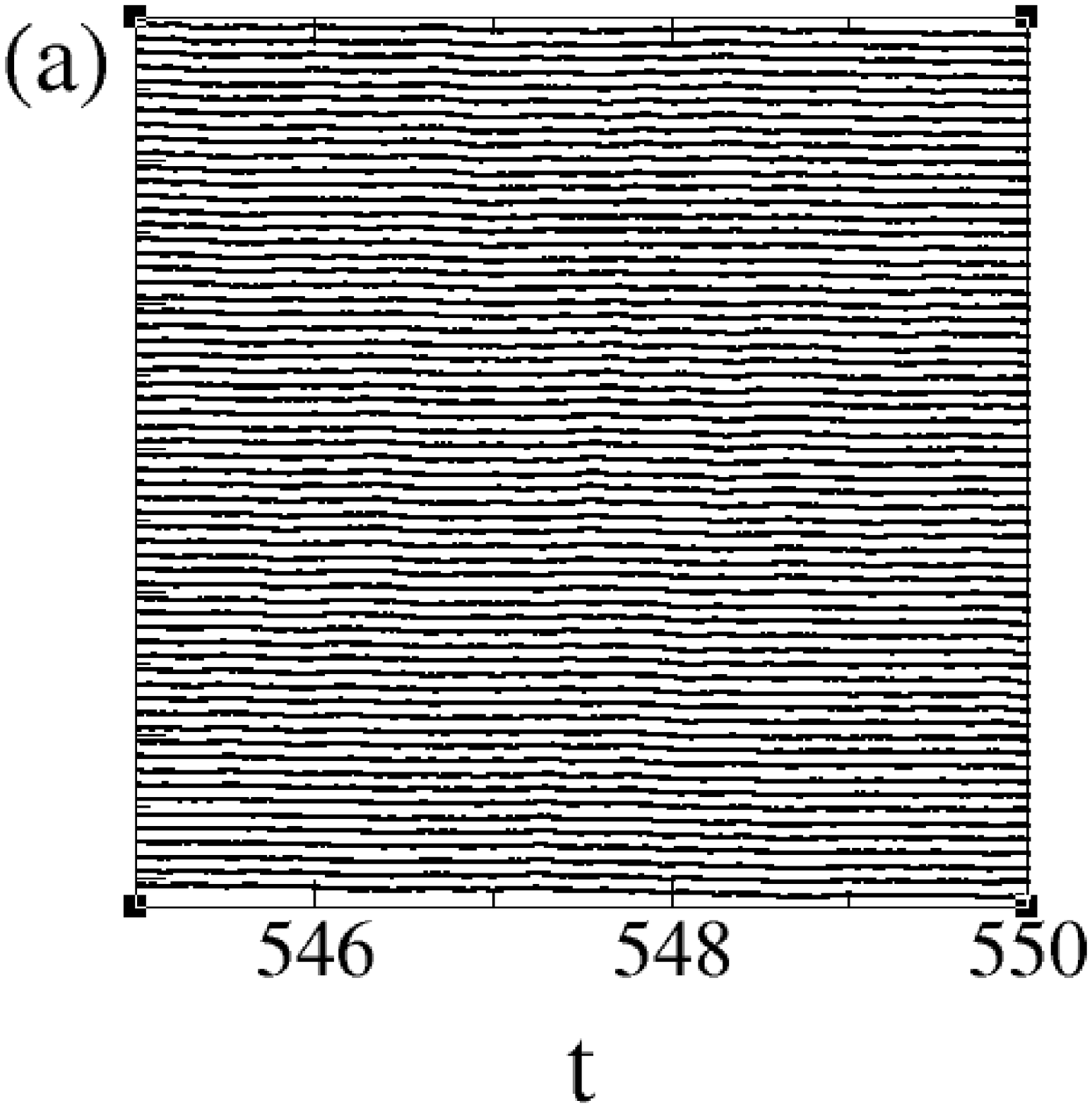}
\hspace{1cm} \includegraphics[clip=true,width=7cm]{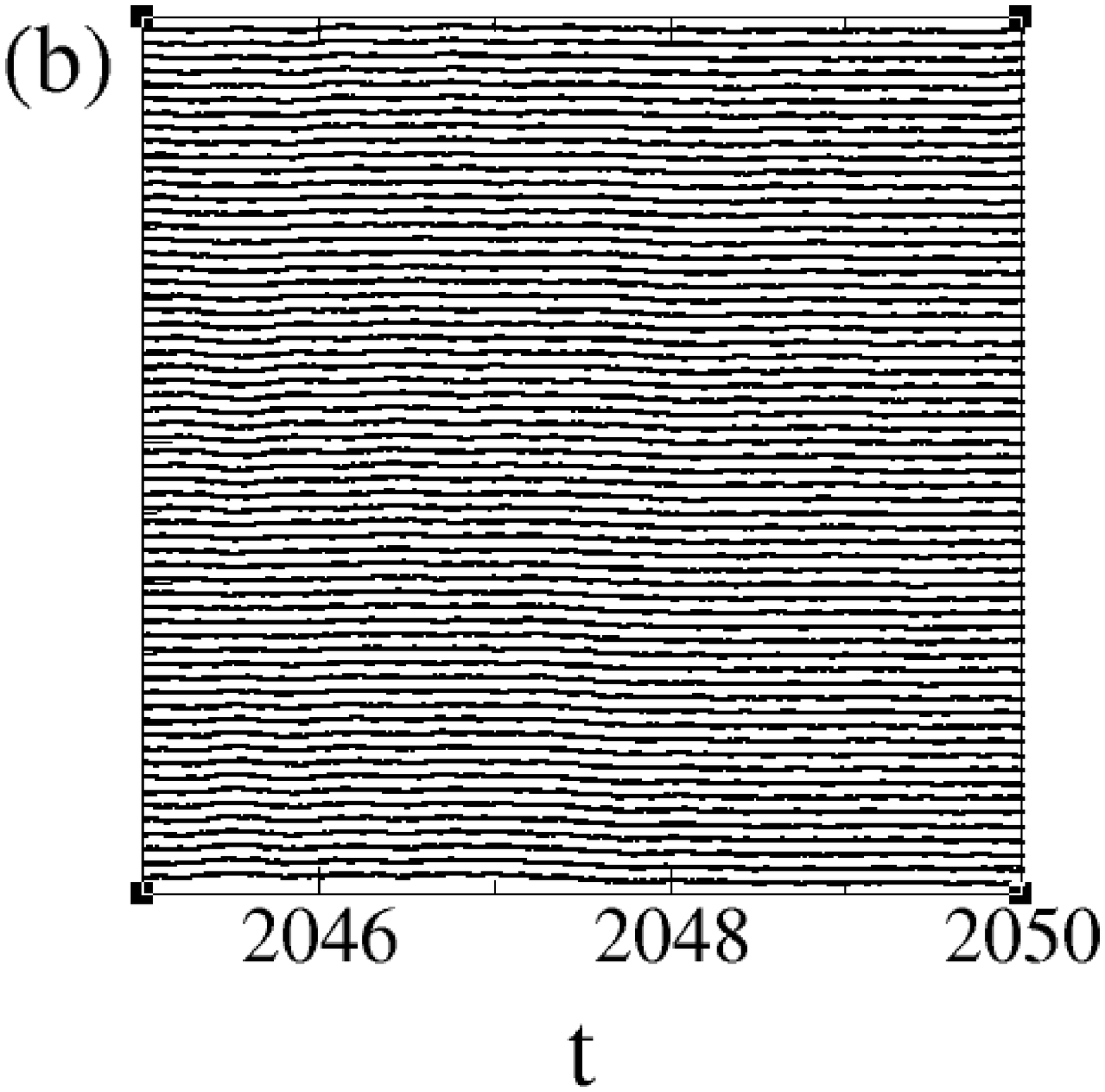}

\includegraphics[clip=true,width=7cm]{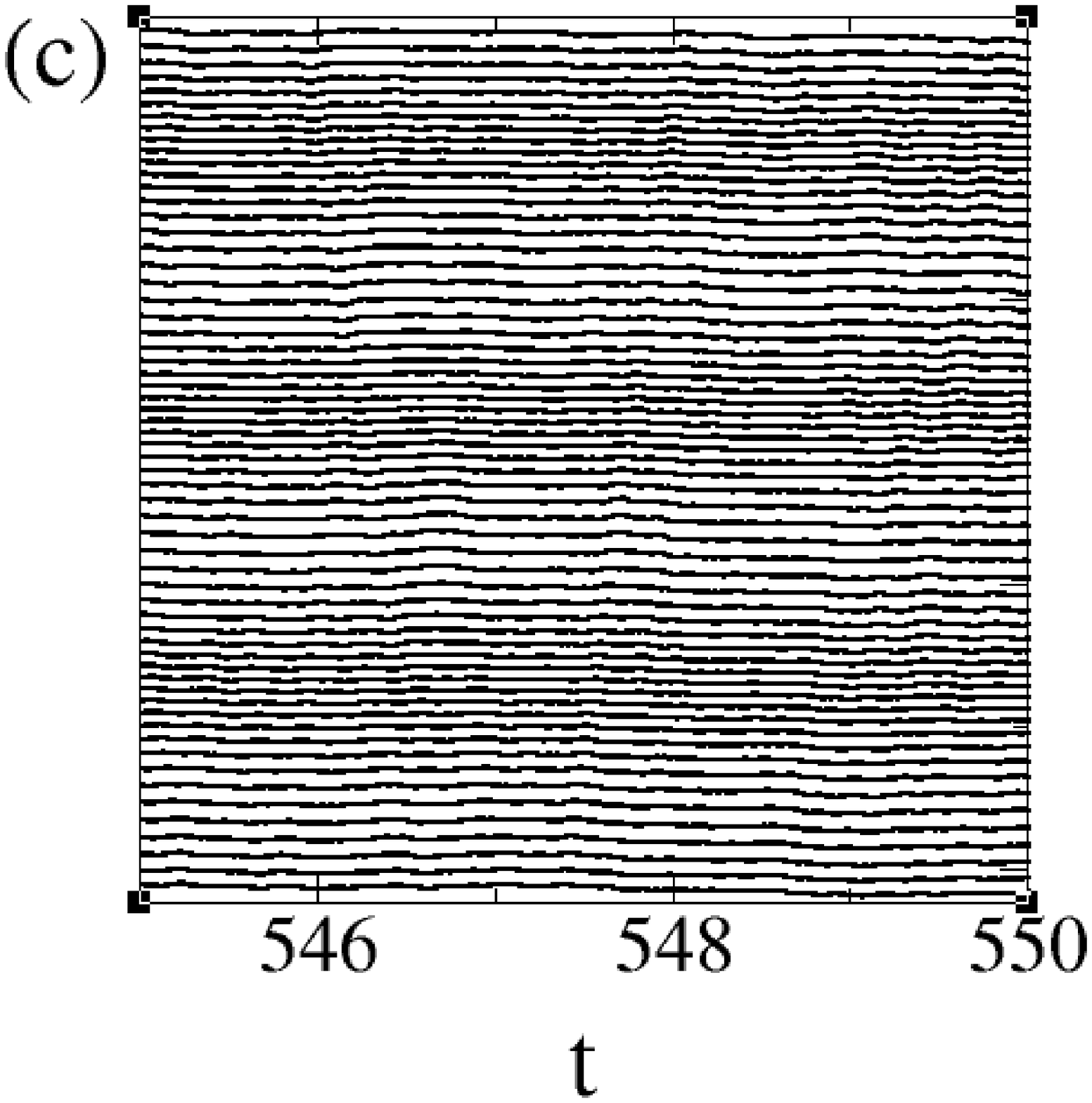}
\hspace{1cm} \includegraphics[clip=true,width=7cm]{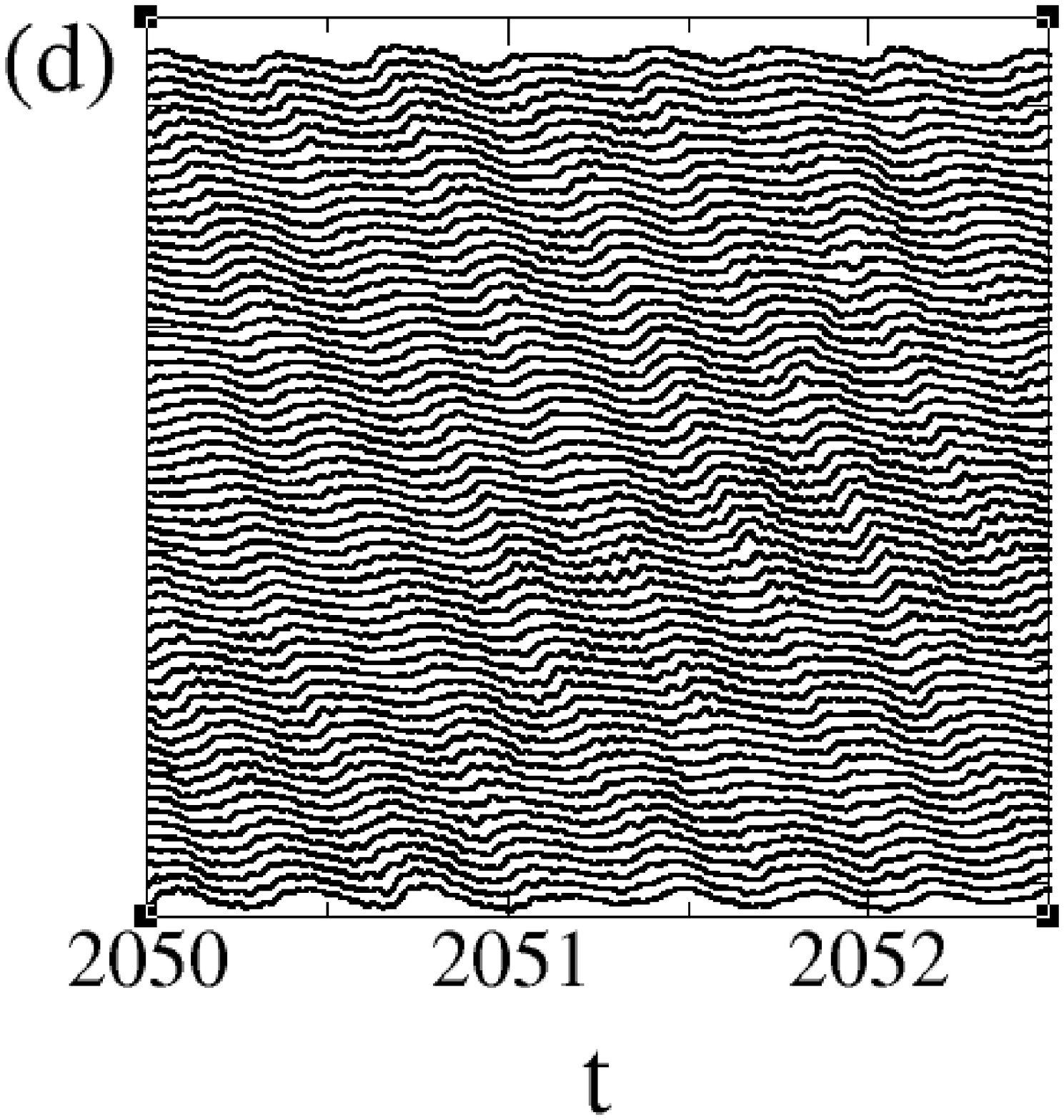}
\caption{Snapshots of the position of the particles as function of time (in
  units of $r_0\sqrt{m/\varepsilon}$), relative to the center of mass position.
  In order to emphasizes visualization of the structural change we have
  diminished the distances between particles. Snaps (a) and (b) correspond to a
  weak friction case, while snaps (c) and (d) correspond to a strong friction
  case; snaps (a) and (c) were taken before the application of force, while (b)
  and (d) were taken after that.}
\label{snap}
\end{figure}

To provide additional support to this interpretation we plot, in
Figs.~\ref{snap}, snapshots of the system as it evolves in time: these snapshots
represent the positions of many particles (a section of the chain) as a function
of time before and after applying the external force, for strong and weak
friction representative cases.

Figures~\ref{snap}(a,b) correspond to a weak friction case, before and after the
force is applied, respectively.  Figures~\ref{snap}(c,d) are the analogous
snapshots for the strong friction case.  Comparing Fig.~\ref{snap}(a) and
Fig.~\ref{snap}(c) there is an obvious difference between the structure of the
chain in equilibrium (without external force), which is due to the constraint
imposed by the substrate potential.  In the strong case (Fig.~\ref{snap}(c)) the
commensuration ratio is such that the system has to significantly rearrange to
accommodate to the substrate potential, which is reflected in the kinks that can
be observed and which yield a stable pattern.  This in turn precludes the system
vibrations in a specific collective mode, thus acting as a barrier for the
propagation of specific frequencies (as if some modes are pinned) generating the
minimum in the density of states of Fig.~\ref{dos}(b). This feature is absent in
the snapshot of Fig.~\ref{snap}(a) which can hardly be distinguished from a free
chain.  So ${\cal D}(\omega)$ looks like a 1D case.  When the external force
sets the system in sliding motion the kink acts as an excitation, forcing the
system to vibrate in what was a forbidden mode for $F=0$.  The frozen mode now
does slide, as can be seen in Fig.~\ref{snap}(d), where the propagation of the
kinks is quite apparent.  Therefore, a strong absorption band appears precisely
where there was a minimum in the force free situation (Fig.~\ref{dos}(b)). The
corresponding weak friction case shows no propagating kink at all and the whole
structure is less perturbed (Fig.~\ref{dos}(a)). However, the excitation of all
the particles is evident, as for a free chain.

In conclusion, it is our understanding that this could explain the large
differences in sliding friction behavior for different commensuration ratios:
the key element is the appearance of a new channel of dissipation (due to large
wavelength kinks) that is responsible for the increased friction in the strong
friction cases.  The kink appears when the adsorbate is near perfect
commensuration with substrate. This channel acts in parallel with the normal
phonon dissipation mechanism.

\begin{figure}[htb]
\center \includegraphics[clip=true,width=7cm]{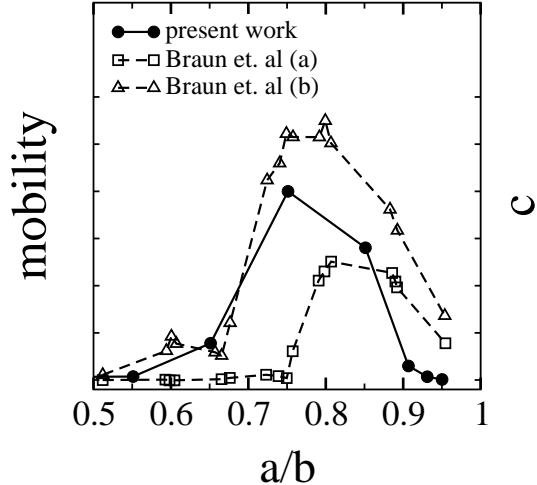}
\caption{Comparison between our results for the $c$ parameter and Braun et
  al. data of mobility as a function of commensuration ratio. Present work
  results are at $k_{B}T=0.01\varepsilon$, while Braun et al. results are at
  $T=0.0025\; eV$ (a) and $T=0.02\; eV$ (b). Curves (a) and (b) of Braun et
  al. are curves (1) and (3) respectively of Fig.2 from Ref.~\cite{Braun96}. The
  coefficient $c$ is in units of $(mr_0^2\varepsilon^3)^{-1/2}$, while mobility
  is adimensional. The scale was adjusted for the comparison between them.}
\label{braun}
\end{figure}

The present results are consistent with those obtained by Braun et
al.~\cite{Braun96} who, using molecular dynamics simulations, studied the
mobility and diffusivity of a generalized Frenkel-Kontorova model taking into
account anharmonic interactions. In Fig.~\ref{braun} we compare our results with
those of Braun et al. From the definition of mobility it is possible to write
\[ B\propto\frac{1}{\eta}\propto\frac{1}{c} \]

\noindent 
where $c$ is the proportionality coefficient of Eq.~(\ref{eq6}).  Notice that
curve (a) has a maximum for the mobility (minimum for the friction coefficient)
in the vicinity of our own maximum, in spite of the different temperatures used
in the simulations.  In the curve labeled (b), where the difference between
temperatures is even larger, the minimum is only slightly shifted to the
right. All of the latter points toward the robustness of our interpretation of
the enhanced phononic friction.

A final comment: Fig.~\ref{coef}(b) shows the $c$ coefficient ---the same that
is plotted in Fig.~\ref{coef}(a)--- but obtained using a lower range of external
forces. If the linear assumption were strictly valid the two graph should be
identical; therefore the linear assumption is not generally valid.  While this
is a very relevant issue, it can be seen that the qualitative general
conclusions presented so far are not affected by this.  
Complementarily, at very low forces~\footnote{Forces in the range
  $0.0001-0.001$ are very low in comparison with previous numerical
  works, but they are still high in comparison with experimental
  setups; i.e. using parameters of Xe, such forces produce velocities
  in the range of $4-20\;m/s$} the region of low phononic friction
between $0.6-0.9$ are in fact of zero phononic friction, thus only
electronic friction should be expected in these cases.

\section{Conclusions}
\label{con}
A phononic friction coefficient calculation for a one dimensional model of an
adsorbate-substrate interface, has been presented.  Using molecular dynamics
simulations we investigated the system, for different substrate/adsorbate
commensuration ratios $a/b$, in order to address two basic questions: first, how
the phononic friction (and therefore the sliding friction) depends on the
commensuration ratio between substrate and adsorbate; and second, whether a
reasonable picture can be found to explain the divergent results between
equivalent calculations in the
literature~\cite{Smith,Persson96,Tomassone97,Liebsch99}.  We believe that both
these goals have been attained since, for specific ranges of $a/b$ values, our
model calculations yield a relatively large friction coefficient as compared
with the intrinsic or electronic friction $\eta_e$.  This picture holds if $a/b$
falls in what we denominate the strong friction region; otherwise, $\eta_{ph}$
can be neglected in comparison with the electronic friction $\eta_e$.

From the technical point of view such sensitive dependence provides a plausible
explanation for the divergent results found in the
literature~\cite{Smith,Persson96,Tomassone97,Liebsch99}.  In fact, a small
rotation of the adsorbate relative to the substrate in 2D or 3D simulations of
sliding friction, can change the commensuration ratio between the surfaces in
such a way as to generate a large change in effective friction, allowing one
group to claim that electronic friction is dominant while the other states the
opposite, both of them on the basis of properly carried out computations.  Our
conclusion is that simulations with much larger systems, and averaged over
different realizations, have to be performed in order to minimize the artifact
imposed by small system simulations or poor averaging.  This suggestion implies
a rather formidable computational challenge, but certainly more feasible than it
was that five or seven year ago.

The present results might appear conflictive when compared with recent results
on the friction of a dimer sliding in a periodic
substrate~\cite{Goncalves04,Goncalves05}. For the latter system the friction due
to vibrations is maximum for a commensuration ratio $a/b = 2/3$. However, such a
friction is due to resonance of the internal oscillation of the dimer that
happens at much higher sliding velocities than the ones used here, since our
purpose is to compare to typical experimental setups.

A final question can now be raised: ``Is the friction measured in the laboratory
sensitive to these subtle changes in the topology, as seen by the sliding
layer?''  One way to check on this would be to try different adsorbates on the
same substrate (for example, comparing Xe, Kr and Ar sliding on Au), or the same
adsorbate sliding over different substrates (Xe on Au or Ag).

\section{Acknowledgments}
EST acknowledges the support of CNPq and CAPES. SG acknowledges
hospitality of the Department of Physics and Astronomy of the
University of New Mexico and support of the National Science
Foundation under grants INT-0336343 and DMR-0097204, during the final
stages of this work.  MK was supported by FONDECyT grant \#1030957.
This work was supported in part by Fundación Andes.

\end{document}